\documentclass[seceqn,dvips]{arxstspdf}
\usepackage{dcolumn}
\usepackage{graphicx}
\usepackage{flushend}
\usepackage{stfloats}

\volume{24}
\issue{3}
\pubyear{2009}
\firstpage{303}
\lastpage{318}
\doi{10.1214/09-STS302}

\makeatletter
\newcolumntype{d}[1]{D{.}{.}{#1}}
\newtheorem{theorem}{Theorem}[section]
\newtheorem{lemma}[theorem]{Lemma}
\makeatother

\begin{document}
\begin{frontmatter}

\title{Model Assessment Tools for a Model False World}
\runtitle{Model Assessment Tools for a Model False World}

\begin{aug}
\author[a]{\fnms{Bruce} \snm{Lindsay}} \and
\author[b]{\fnms{Jiawei}
\snm{Liu}\ead[label=e2,text=matjxl@ langate.gsu.edu]{matjxl@langate.gsu.edu}\corref{}}
\runauthor{B. Lindsay and J. Liu}

\affiliation{Pennsylvania State University and Georgia State University}

\address[a]{Bruce Lindsay is Willaman Professor, Department of Statistics,
Pennsylvania State University.}
\address[b]{Jiawei Liu is Assistant Professor, Department of Mathematics
and Statistics, Georgia State University \printead{e2}.}

\end{aug}

\begin{abstract}
A standard goal of model evaluation and selection is to find a model
that approximates the truth well while at the same time is as
parsimonious as possible. In this paper we emphasize the point of
view that the models under consideration are almost always false, if
viewed realistically, and so we should analyze model adequacy from
that point of view. We investigate this issue in large samples by
looking at a model credibility index, which is designed to serve as a
one-number summary measure of model adequacy. We define the index to
be the maximum sample size at which samples from the model and those
from the true data generating mechanism are nearly
indistinguishable. We use standard notions from hypothesis testing
to make this definition precise. We use data subsampling to estimate
the index. We show that the definition leads us to some new ways of
viewing models as flawed but useful. The concept is an extension of
the work of Davies [\textit{Statist. Neerlandica} \textbf{49} (1995) 185--245].
\end{abstract}


\end{frontmatter}

\section{Introduction}\label{sec1}

Our starting point is the famous quotation of\break G.~E.~P. Box:

\begin{quote}
All models are wrong, but some are useful (\citeyear{box1976}).
\end{quote}

In this article we will take as our initial premise that ``All
models are wrong,'' and see where it leads us. A consequence of model
falseness is that for every data generating mechanism there exists
a sample size at which the model failure will become obvious.



Our second premise is that there are occasions when one will want to use, in
some fashion, a model that is clearly false, provided that it provides a
parsimonious and powerful description of the generating mechanism. Here we
wish to emphasize that we are interested in description, not prediction, as
there is a smaller advantage to simplicity when the overarching goal is
accurate prediction.

In order to explore this question, the key assumption of this paper
will be that the sample size under which the data is collected, say,
$n$, is sufficiently large that many of the models under
investigation are clearly false. This would seem to be a reasonable
assumption in the modern data-mining environment. Just the same, we
wish to measure the quality of their approximation to the true data
generating mechanism to see which ones most economically capture its
main features. Later in this paper we will use subsampling from the
data as a means of replicating the true data generating mechanism.

It is important to our theme that we are seeking to measure
attributes that are completely unrelated to the value of $n$ that
generated the data at hand. We emphasize this because the standard
tools for model assessment are highly $n$-dependent. For example,
hypothesis testing has played a prominent role in the assessment of
the models since the development of Pearson's chi-squared statistic.
Unfortunately, it is based on the false premise that the model is
correct, and so for a large enough sample size, we are doomed to
reject any fixed model. That is, if we view these tests as answers
to the question: ``Is this model useful?,'' then what we mean by
usefulness is clearly related to not just the quality of the model,
but also the size of the sample that was used in its assessment. So
hypothesis testing does not meet our need directly.

In our approach we use testing methodology but in an inverted
fashion. We treat the null hypothesis as being false, and ask
questions about the power of the test statistic as a function of its
sample size. We define our new index, called the model credibility
index, as the sample size needed to obtain a desirable power.
Although the point of view is not new that the power of a test
depends on the sample size, it is a novel idea to propose the sample
size as a model evaluation index.

Other standard risk analyses, the basis for AIC, Mallow's $C_p$ and
other methods are $n$-dependent because the goal there is to assess
the quality of prediction using the fitted model. These criteria for
model selection depend not just on the model itself, but also on the
quality of the parameter estimation, which in turn depends on~$n$.


We hope that our new methods will be thought-provoking because they
involve only standard tools of testing and risk assessment, so they
could be readily understood (and constructed) by any statistician.

Just the same, we think that our work presents a challenge to the
standard statistical train of thought. Statisticians are quite
accustomed to taking the\break ``model true'' point of view.  After all, we
have a huge box of statistical tools that are based on the
assumption. This can make it hard for statisticians to maintain
consistently a ``model false, but maybe useful'' point of view.


For example, suppose we have a random sample $X_1,X_2,\ldots,X_n$ with distribution $\tau$. In traditional model building much
is made of the idea of consistency, in the sense of finding the true
distribution $\tau$ based on the assumption it lies within some
narrow set of models. However, this true distribution is very likely
to be much too complex to be useful, especially if we consider the
discretization, rounding, misrecording and measurement errors
incumbent in real data. (For example, see the discussion of
Ghosh and Samanta, \citeyear{GohsSama2001}, page~1140.) For the duration of this article, at
least, we ask the reader to believe in model-falseness, and further
believe that usefulness is not necessarily tied to consistency.

In the next subsection we give an informal introduction to our
methodology. This will be followed by a more detailed look at the
contents of the paper.

\subsection{Introducing Credibility Indices}

Davies (\citeyear{Davi2002}) gave the following definition:
\begin{quote}
A probability model $P_{\theta}$ is an adequate approximation for
the data set $(x_1, \ldots, x_n)$ if ``typical'' samples
$(X_1(\theta), \ldots, X_n(\theta))$ of size $n$ generated using
$P_{\theta}$ ``look like'' the real data set $(x_1, \ldots, x_n)$.
\end{quote}

This is clearly an $n$-dependent assessment, but it captures what we
consider an important aspect of a good model---that it is good at
creating data similar to the observed data.

To illustrate our thinking, let us start with the most prominent
statistical assumption, that the data is normally distributed.
Surely we might believe that no data is exactly normal in
distribution, but that it is often useful and plausible to assume
so.

Berkson (\citeyear{berk1938}) described the paradox that a good\-ness-of-fit test may
become embarrassingly powerful whenever the data are extensive:

\begin{quote}
I believe that an observant statistician who has had any
considerable experience with applying the chi-square test repeatedly
will agree with my statement that, as a matter of observation, when
the numbers in the data are quite large, the $P$'s tend to come out
small. Having observed this, and on reflection, I make the following
dogmatic statement, referring for illustration to the normal curve:
``If the normal curve is fitted to a body of data representing any
real observations whatever of quantities in the physical world, then
if the number of observations is extremely large---for instance, on
the order of 200,000---the chi-square $P$ will be small beyond any
usual limit of significance.''

If this be so, then we have something here that is apt to trouble
the conscience of a reflective statistician using the chi-square
test. For I suppose it would be agreed by statisticians that a large
sample is always better than a small sample. If, then, we know in
advance the $P$ that will result from an application of a chi-square
test to a large sample there would seem to be no use in doing it on
a smaller one. But since the result of the former test is known, it
is no test at all!
\end{quote}

As a response, Hodges and Lehmann (\citeyear{hodglehm1954}) suggested that the
difficulty could be avoided by making distinction between
``statistical significance'' and ``practical significance'' in the
formulation of the problem. The idea was to construct a larger
hypothesis $H_1$ of distributions about the null $H_0$, representing
distributions that are close enough to $H_0$ so that the difference
is deemed not practically significant with the data at hand. If one
let $H_1$ play the role of the null hypothesis, then if the true
distribution is an element of $H_1$, then one might still wish to
use the model $H_0$. Liu and Lindsay (\citeyear{Liulind2009}) expanded upon this idea,
but still found difficulty in creating a reasonable set $H_1$ having
a simple interpretation.



\begin{figure}[b]

\includegraphics{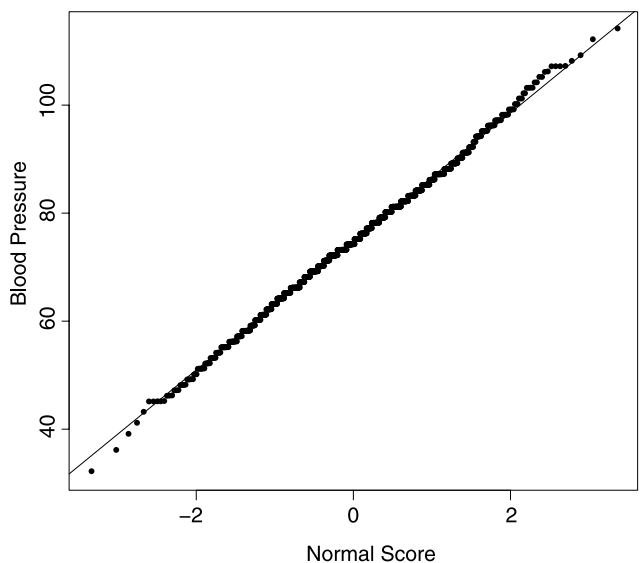}

\caption{QQ plot of the Blood Pressure data of 1239 females.}\label{fg:sat}
\end{figure}

Conducting a goodness-of-fit test involves two\break choices: the test and
the significance level $\alpha$. Given an alternative, there is a
resulting type II error $\beta$. We start our development by showing
how one can invert goodness-of-fit testing to develop a new measure
of model failure. To help fix the idea, we use the following
example. The full data set consists of the diastolic and systolic
blood pressure data of 10,529 persons aged from 35 to 84. We take
only the 1239 normal females as our data to be analyzed, because
the blood pressures of the full sample would likely be better
modeled as a mixture of normals. The original data was obtained from
the Clinical Trials Research Unit (CTRU) of New Zealand. Central
limit theory suggests that such data might be rather normal in
distribution. After looking at the QQ plot Figure~\ref{fg:sat},
where there is little deviation from a straight line except at
tails, we think many statisticians would be happy using a normal
model for such data.

On the other hand, suppose we use the Kolmogorov--Smirnov
goodness-of-fit test to test the normality assumption. The test
statistic is the greatest absolute vertical distance between the
empirical distribution function of blood pressures and the
hypothetical normal distribution function, evaluated on the 1239
sample values. The parameters of the normal distribution are
estimated from the sample. Normality is strongly rejected ($p$-value
$=0.0016$), a fact which we might attribute to the large sample size
($n=1239$). That is, at such a sample size, we have power against
what appear to be very small deviations from normality. In this
example, the normality is rejected although data looks quite normal
at the center.

How can we say this data is very well described by a normal model
without saying it is \textit{exactly }normal? Here is one way to use
statistical testing to answer the question.

One starts with a goodness-of-fit test method that has desirable
operating characteristics. That is, it should be sensitive to
important model failures (alternatives) but insensitive to trivial
model failures. We discuss this choice in the next subsection.

\begin{figure}[b]

\includegraphics{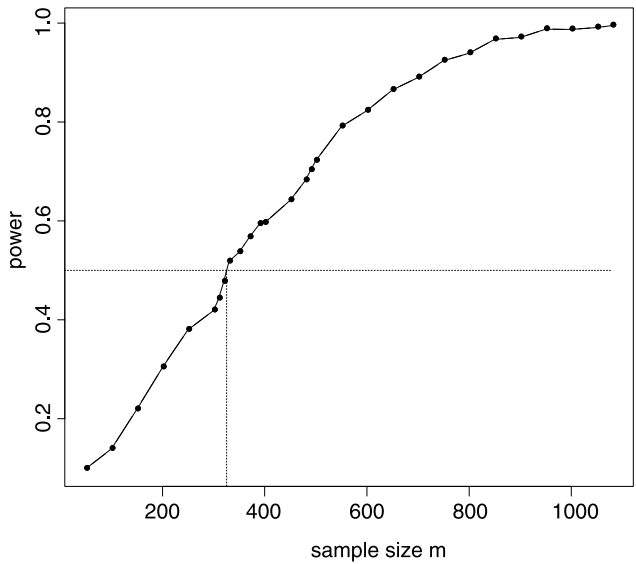}

\caption{Plot of test power vs. sample size.} \label{power curve}
\end{figure}

Given a true probability generating mechanism $\tau$, that is not
in the model, and a size $\alpha$ test procedure
$I   \{T_m(X_1,\ldots, X_m)>c_m  \}$, one  can define the power curve
$\beta_{\tau}(m) =P_{\tau} \{T_m(X_1,\ldots, X_m)>c_m  \}$.
See Figure~\ref{power curve} for such a plot based on the blood
pressure data. Here $\tau$ is the empirical distribution of the full
data set, the test is the Kolmogrov--Smirnov test for normality with
$\alpha=0.05$. As a simple number summary of such a plot, we define
the \textit{maximum credible sample size} of the postulated model
(here the normal model
in the blood pressure population) to be that sample size
$N^{\ast }=N^{\ast }(\tau, \mathcal{M} )$ at which we would reject
the model $\mathcal{M}$ 50\% of time based on a size $\alpha$ ($<$0.5)
goodness-of-fit test. We will also call $N^{\ast }$ the \textit{model
credibility index.} More generally, one could define $N^*_{\beta}$
as the sample size needed to attain power $\beta$, in which case the
index $N^*$ is $N^*_{0.5}$.

Although one might choose other summaries of the power curve, such
as $(N^*_{0.25}, N^*_{0.75})$, we find $N^*$ to be a natural
summary. It also creates certain asymptotic simplifications.

\begin{table}[b]
\caption{$m$ at  various test sizes and power levels for blood pressure data}\label{normal tb}
\begin{tabular*}{\columnwidth}{@{\extracolsep{\fill}}lccc@{}}
\hline
& \multicolumn{3}{c@{}}{\textbf{Test size}} \\
 \cline{2-4} \\[-6pt]
\textbf{Power} $\bolds{ \beta_{\tau}(m)}$ &  $\bolds{\alpha=0.1}$ &  $\bolds{\alpha=0.05}$ &  $\bolds{\alpha=0.01}$ \\
\hline
0.3 & 115 & 200 & \phantom{0}410 \\
0.5 & 225 & 315 & \phantom{0}600 \\
0.7 & 360 & 490 & \phantom{0}795 \\
0.9 & 540 & 695 & 1050 \\
\hline
\end{tabular*}
\end{table}

If the model is actually correct, then $N^{\ast }=\infty$. However,
if the model is false, there is some finite sample size at which the
power would reach 0.50. Different tests will have different power
curves that in turn reveal different inadequacies of the model.

In Figure~\ref{power curve} we assumed that the true distribution
$\tau$ is random sampling from our set of 1239 scores, and we
determined $\beta(m)$ by simulation. That is, we bootstrapped
repeated samples of various hypothetical sizes $m$ from the 1239
blood pressure values and repeatedly conducted the
Kolmogorov--Smirnov test until we found the $m$ that gave power 0.5.
For example, in our example we found when $m=315$, the normality
assumption was rejected by the Kolmogorov test approximately 50\% of
the time $(499/1000)$.

The choice of test size is also arbitrary. Table~\ref{normal tb}
shows the estimated sample size $m$ when obtaining various power
$\beta_{\tau}(m)$ at a different testing significance. The monotone
pattern in the table indicates that one would need a larger sample
size in order to obtain more testing power at a higher test size.

Based on this analysis, it is clear that it would be very hard to
detect non-normality in samples of size 100 from this true
distribution ($\beta(100)=0.13$). To put this another way, the
samples of size 100 must ``look'' very much like samples from a
normal distribution, and so one might say that normality is a good
descriptor of the sampling mechanism
at this sample size. Indeed,
this descriptive power holds till the sample size approaches 315,
when the distinction between normal samples and data mechanism
samples must start to become more obvious.




\subsection{Role of Test Statistics} \label{test stat role}

What kind of index is $N^{\ast }$, in a mathematical sense? As we
will see later, in a detailed analysis of some standard test
statistics, it is inversely proportional to the squared distance
measure that was used to construct the test statistic.

This makes it quite clear that the value of the model credibility
index $N^{\ast } $ depends strongly on the test statistic that is
being used. If we wish $N^{\ast }$ to reflect usefulness of the
model, then the test statistic must be sensitive to those model
failures which we consider most important. Thus, the choice of the
test must reflect our statistical purposes, as well as which models
we consider to be competitors. For example, if we would consider a
$t$-distribution a useful alternative description, having a test
sensitivity to tail probabilities would be desirable, say,
Anderson--Darling.


The Kolmogorov--Smirnov test is a test of normality for large samples.
One of its limitation is that it is more sensitive to deviations in
the center rather than in the tails. In the blood pressure example,
at least the center of data is quite normal (Figure~\ref{fg:sat}).
If one is interested in the tail regions, then one should use other
tests that are more sensitive to tails. More generally,
Claeskens and Hjort (\citeyear{ClaskHjort2003}) develop model selection tools which can focus on
specific aspects of lack of fit.



\begin{figure*}

\includegraphics{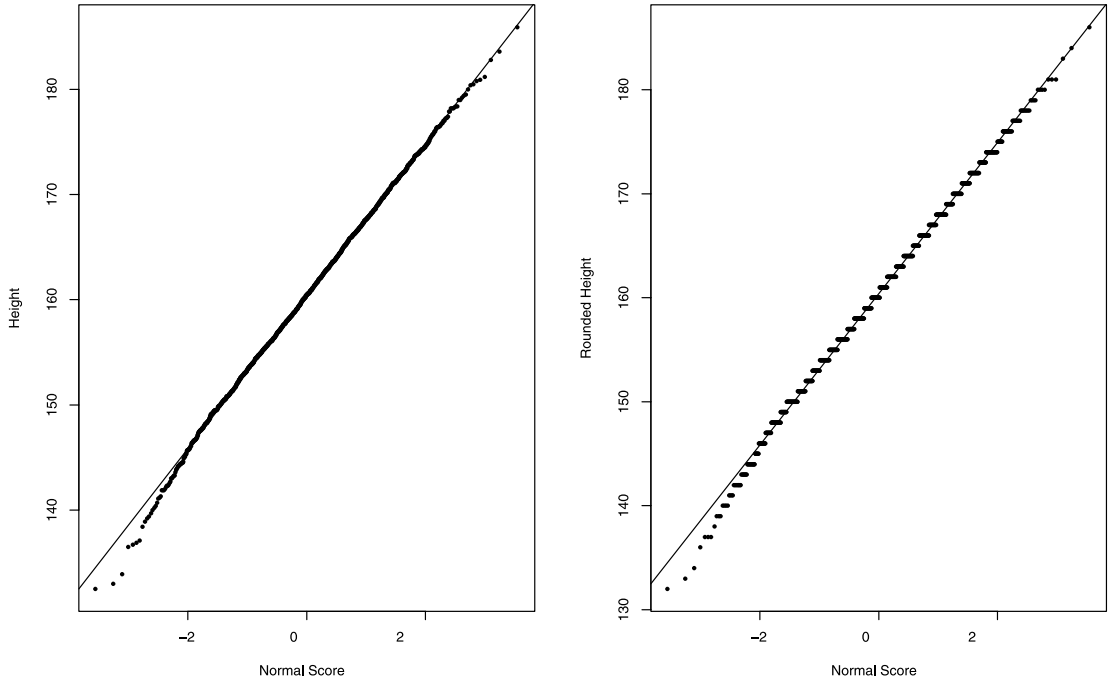}

\caption{QQ plot of the heights of 2603 female adults, both original and rounded data.} \label{fg:height}
\end{figure*}

While trying out other data sets to use in this paper, we examined
another data set with heights of 2603 female adults from the data
surveys and collection systems of the Centers for Disease Control and
Prevention (NHANES, 1999--2000). The Kolmogorov--Smirnov test for
normality of this set gave a $p$-value greater than 0.10. Although
this data set didn't meet Berkson's criterion of 200,000, it was
even more normal than the blood pressure set. See Figure~\ref{fg:height}.
We found another interesting thing for this heights
data. The original data is coded in centimeters with one decimal
accuracy. However, when we rounded the data to integer values, the
$p$-value of the Kolmogorov--Smirnov test became $0.000$, leading to a
rejection of normality. This illustrates that the Kolmogorov--Smirnov
test is sensitive to data coding.

The Shapiro--Wilks $W$-statistic (\citeyear{shapwilk1965}) is a well-known goodness-of-fit
test for the normal distribution. It is attractive because it has a
simple, graphical interpretation: one can think of it as the
correlation between given data and their corresponding normal
scores. The Shapiro--Wilks test has good power properties across a wide
range of alternative distributions in comparison with other goodness-of-fit tests (Shapiro, Wilk and Chen, \citeyear{shapwilkchen1968}).

For the blood pressure data, normality is also rejected by
the Shapiro--Wilks $W$-statistic ($p$-value${}=0.0043$). The credibility index is
$N^{\ast }=220$ for the Shapiro test.

The chi-square test, introduced by Pearson in 1900, is the oldest
and best known goodness-of-fit test. The idea is to reduce the
goodness-of-fit problem to a multinomial setting by grouping data
and comparing cell counts. Chi-squared tests can be applied to any
type of variable: continuous, discrete or a combination of these.
However, grouping the data sacrifices information, especially if the
underlying variable is continuous. For the blood pressure data,
normality is rejected by the chi-squared test with $p$-value${}={}$0.0000;
and the credibility index is $N^{\ast }=240$.

In comparing these credibility indices, we recall that---even though
$N^{\ast }$ has a natural sample size inter\-pretation---it is
$\sqrt{N^{\ast }}$ that is the more statistically meaningful
quantity, as it reflects the standard deviation scale of
uncertainty. (This in turn arises, mathematically, because $N^{\ast}$
is inversely proportional to the squared distance, making its root
inversely proportional to the distance.) For these tests, the root
indices were $\sqrt{315}=17.75$, $\sqrt{220}=14.83$, and
$\sqrt{240}=15.49$, very similar values, albeit measures of
different model fit features.

How might one use the $N^*$-index? Certainly in any particular data
set $N^*=315$ has its own direct statistical interpretation. And one
can use simulation methodology to obtain a better feel for the
magnitude of $N^*=315$, as we do in Section~\ref{1 and 2 sample}. More generally,
given a specific testing method and type of data set, one could use
the $N^*$-values to address the question as to which data set is a
better fit to the model and quantify the differences. However, the
greatest strength of this methodology is that it creates a universal
tool that transcends particular data types and particular testing
methods. That in turn raises questions as to whether it is possible
to compare $N^*$-values across different settings in a reasonable
way. In particular, one might ask whether an $N^*$-value is large or
small given the number of parameters included in the model. This
last question we defer to future research.

\subsection{Estimating $N^*$}

To this point, we have treated $N^*$ as a population quantity, where
the population in our example is a large data set. As such, there is
only simulation error in our bootstrap estimation. Inference about
$N^*$ when the large data set is itself treated as a sample of size
$n$ from a yet large population, so $\tau$ is unknown, creates some
challenging inference problems. One can, as before, estimate the
power curve $\beta_{\tau}(m)$ by averaging over bootstrap samples of
size $m$, but now the estimator is not unbiased for
$\beta_{\tau}(m)$ unless we use sampling without replacement, a
method we will simply call subsampling (see Politis, Romano and Wolf, \citeyear{poliromawolf1999}).

The subsampling framework gives us several tools to tackle
inferential questions. In a later section we will show that we have
consistent and asymptotically normal estimation of $\beta_{\tau}(m)$
when $m$ is fixed and $n \rightarrow \infty$. However, in a more
realistic scenario in which the sampling fraction $\phi=m/n$ is
fixed as $n \rightarrow \infty$, the inverse ratio $\phi^{-1} =n/m$
is shown to be an important measure of the quality of $N^*$
inference. When $\phi^{-1}$ is small, say, 10 or less, then the estimator
of $\beta_{\tau}(m)$ has considerable uncertainty.

\subsection{Our Contents}

We have now introduced a measure of the credibility of a model which
depends on the hypothesis testing methodology, but it comes with a
new interpretation. Note that it is a characteristic of the model,
the test statistic and the data generating mechanism, but not the
\textit{de facto} sample size $n$ used to estimate it. It is a highly
portable statistic, as one can use it in any context where there is
a known goodness-of-fit procedure. However, it is also clear that it
can only be estimated well when the \textit{de facto} sample size is
large enough to make the model in question clearly false.

In this paper we start by discussing how the work of Davies inspired
our approach in Section~\ref{sec note}, and reviewing briefly other related
literature. We then formally define the model credibility index in
Section~\ref{sec3}. There we also expand upon the normal example so as to
compare numerically two-sample and one-sample testing approaches and
to compare bootstrapping and subsampling as methods to compute
$N^*$.

In Section~\ref{sec4} we explore the asymptotic properties of the power
estimators associated with the model credibility index. We then in
Section~\ref{sec5} examine the structure of the model credibility index in
greater detail in the context of likelihood ratio testing in
categorical models. We will show how these indices are closely
related to Kullback--Leibler discrepancy measures, and give some
further numerical examples.
Section~\ref{sec6} concludes the paper and
proposes topics worthy of further investigation.

\section{Background}\label{sec note}

In this section we will review some related work on the conceptual
difficulty involved in using models while assuming they are false.

\subsection{Distance-Based Indices of Fit}

A more standard approach to model-false analysis would be to
characterize model fitness by choosing a suitable distance measure,
then doing inference on the distance between the true distribution
and the model.

In 1954 Hodges and Lehmann proposed using tolerance zones around the
null hypothesis. They constructed $H_1$ as a set of distributions
whose distance to $H_0$ doesn't exceed a specified bound $c$ under a
distance measurement. Hodges and Lehmann's analysis was in the
context of the chi-squared goodness-of-fit test. They used a
weighted Euclidean distance as the distance from a model element to
the truth. The usual chi-squared distance is included by choosing
appropriate weights.

Hodges and Lehmann didn't give a detailed discussion on how one
should choose $c$. They mentioned that the specification of $c$
would ``present problems similar to those encountered in choosing
the alternative at which specified power is to be obtained.'' This
quoted statement presents some difficulties in its interpretation
and implementation.

Liu and Lindsay (\citeyear{Liulind2009}) expanded on  this tubular model idea, but
used two different distances, likelihood for the test statistic and
Kullback--Leibler for the tube hypothesis. Their tubular model
consisted of all multinomial distributions lying within a
distance-based neighborhood of the parametric model of interest. The
distance between the true multinomial distribution and the
parametric model was used as the index of fit. Liu and Lindsay
developed a likelihood ratio test (LRT) procedure for testing the
magnitude of the index.

Goutis and Robert (\citeyear{GoutRobe1998}) proposed a Bayesian approach for the model
selection problem based on the likelihood deviation between two
nested models, called the full and restricted models. The full model
space was considered to contain the true distribution. The Bayesian
approach was implemented by specifying a prior distribution in the
full model, possibly an improper prior. Each prior distribution was
projected onto the restricted model space and the corresponding
minimum distance measure was computed. Therefore, the posterior
distribution of the distance from a prior distribution to the
restricted model can be derived. Bayesian inference was made on the
restricted model based on the posterior distribution. For example,
one criterion was to reject the restricted model if the posterior
probability that the distance was less than a certain bound $c$ was
small enough. Other aspects of the posterior distribution could be
considered as the testing criteria. When one doesn't have a strong
prior belief, several priors could be used to assess the distance
between models. The sensitivity of the inference to the priors could
be used as a factor in making the model choice.

Dette and Munk (\citeyear{DettMunk2003}) used the Euclidean distance in the problem of
testing for a parametric form hypothesis in regression. They assumed
that the true model was an unknown nonparametric regression
function. Goodness of fit was measured by the Euclidean distance
between the unknown true regression function and the parametric
model.

Dette and Munk first estimated the Euclidean distance under the null
hypothesis. To obtain the distance under the alternative, the
classical concept of analysis of variance was generalized to the
nonparametric setting. Their goodness-of-fit statistic measure could
be interpreted as the difference between variance estimators under
the null model and the nonparametric model.

The challenge one faces with all approaches that use distances
directly, such as those described above, is that it is very
difficult to give statistically meaningful interpretations to the
numerical values of the distance. The credibility indices we have
explored here are, in essence, reciprocals of such distances.
However, we believe that they are easier to interpret, as they
measure the ability of the model to describe samples of various
sizes. They are also more universal, having meaning across a wide
range of settings.

\subsection{Davies}

In our search for a reasonable way to measure how well a model
describes a data generating mechanism, we came across the work of
Davies.

Davies (\citeyear{Davi1995}) proposed the idea of judging model adequacy using the
concept of data feature. The basic idea is that if samples that are
simulated from the model are largely indistinguishable from the real
data, then the model should be regarded as adequate. A similar idea
is expressed in Donoho (\citeyear{Dono1988}) via the following statement:
``No distribution which produces samples very much like those actually
seen should be ruled out a priori.''

Davies' formal theory of data features is very similar to hypothesis
testing for goodness of fit, with the test statistics being designed
to assess whether the data had the same features as a sample from
the model. In common with testing theory (but contrary to us), he
measures the adequacy of models from the null-centric convention
(i.e., that the model is correct) and does so at the \textit{de facto}
sample size.

Another distinction from our approach is that rather than using
model-based one-sample test statistics, he would use a nonparametric
two-sample test to compare the data not with the model, but with
samples from the model. This has the conceptual advantage of being a
direct answer to the question ``Does this data look like a typical
sample from the model?''

The disadvantage to this approach is that it limits the number of
testing procedures available for model assessment. We believe that a
one-sample test is addressing the right question, but it does have
more power because it removes sampling uncertainty. An example in
Section~\ref{1 and 2 sample}  shows that there would be a
substantial change in magnitude of $N^*$ if we used a two-sample
approach.

\section{Credibility Index}\label{sec3}

\subsection{The Formal Definition}

In constructing these indices, we have used the conventional test
size $\alpha =0.05$. For a given test, we let $N^{\ast }= N^{\ast}(\bolds{\tau },\mathcal{M})$ be the value of $n$ that gives
this test power $0.5$ at true distribution $\tau $, when the model
is $\mathcal{M}$. Any test that is consistent for every alternative
hypothesis (i.e., an omnibus test of fit) will give a finite
$N^{\ast }$ under the false model assumption.

The choice of $\alpha $ here seems like an arbitrary element, but we
will see later that it plays a minor role in the comparison of
$N^*$-values. The choice of power 0.5 is also somewhat arbitrary,
but there are two strong reasons behind this choice. First, there is
the intuitive appeal of the idea that the model decision is 50$/$50 at
this point and so the decision is ``up for grabs.'' The index is the
middle value of the power curve and so provides a natural one number
summary (e.g., Figure~\ref{power curve}). Second, this value of the power greatly
facilitates the asymptotic analysis, as we will soon see.

In an intuitive sense, the model credibility index $N^{\ast}(\bolds{\tau },\mathcal{M})$ operates reciprocally to distance
in the following sense. When a true distribution $\bolds{\tau}$ is moved closer to the model, so the distance is reduced, the
sample size index should increase because a larger sample size $n$
would be needed for discrimination between $\tau $ and $M$.
Typically goodness-of-fit test statistics are based on distance
measures; in these cases the reciprocal connection can be made more
precise, as we will soon see.

\subsection{Determination of $N^{\ast }$}\label{sec:boots}

One attractive thing about the testing index $N^*$ is that it admits
an elementary subsampling estimation. This could be carried out in a
typical IID setting as follows.

Given a target size $\alpha $ and a data set $x_{1},\ldots,x_{n}$,
suppose one would ordinarily conduct the goodness-of-fit test of the
model based on an asymptotic critical value. One could then estimate
$N^{\ast }$ for this test procedure by conducting a nonparametric
bootstrap simulation using various sample sizes $m$ to
estimate the power $\beta_{\tau}(m)$, the goal being to find the value of $%
m$ such that $\beta_{\tau}(m)= 0.5$. If we
let the symbol $\hat{F}$ represent the empirical distribution, we are treating
$\hat{F}$ as $\tau$, and calculating $\hat{N}^{\ast }=N^{\ast }(\hat{F},\mathcal{M})$.
Now assuming the model $\mathcal{M}$ does not include the
empirical distribution $\hat{F}$, the bootstrap sampling
distribution is under the alternative, and so the rejection
probability, which is the power of the test, should increase in $m$.
The female blood pressure example in Section~\ref{sec1} is an example of the
bootstrap determination of $N^*$.

As we will explain later, there are good reasons to use sampling
without replacement (``subsampling'') instead of with replacement
(``bootstrapping''). In subsampling the largest possible value of $m$
is $n$, and the resulting estimated power $\hat{\beta}(n)$ is 1, if
the test rejects, and $0$, if the test accepts. This reflects our
lack of knowledge (in the model false world) about the model's
capacity to explain future samples of size $n$ or larger.

To carry out a subsampling or a bootstrap determination of $N^{\ast
}$, one needs to define an efficient algorithm so as to minimize
computation time. Obviously, sensible interpolation methods should
be used. Moreover, it would be nice to have a good starting value
based on asymptotic approximations. See Section~\ref{thetatau} for more on this
issue.

\subsection{One-Sample and Two-Sample Indices} \label{1 and 2 sample}

In this section we use a particular simulation model to compare
different ways of computing $N^*$. We start by comparing one- and
two- sample credibility indices. In this process we also learn
something more about how to interpret the magnitude of a model
credibility index.

Suppose we draw two samples of size $m$, say, one each from a normal
and a logistic distribution, where the parameters are chosen to make
the distributions as similar as possible. We could measure their
similarity by using a two-sample test to see if the samples are
detectably different. Doing this repeatedly gives us the power of
the two-sample test between the two distributions.

We did this using the two-sample Kolmogorov--Smirnov test, using 1000
samples for each $m$. Table~\ref{logistic} lists the number of
rejections for various sample sizes.
\begin{table}
\tablewidth=6.5cm
\caption{Power of the Kolmogorov--Smirnov test (two-sample method) to
detect the difference between normal and logistic distributions at selected sample sizes}\label{logistic}
\begin{tabular*}{\tablewidth}{@{\extracolsep{\fill}}rc@{}}
\hline
\multicolumn{1}{@{}l}{$\bolds{m}$} & \textbf{Rejection proportion} \\
\hline
100                 & 0.044 \\
500                 & 0.116 \\
1000                & 0.169 \\
2000                & 0.361 \\
$\hat{N}^*=2650$    & 0.513 \\
4000                & 0.768 \\
6000                & 0.907 \\
\hline
\end{tabular*}
\end{table}

Suppose we let the \textit{model credibility index} $N^{\ast }$ be the
value of $n$ that gives power 0.50. In this example, $N^* \approx
2650$. We found it quite striking that the normal and logistic
models would be so poorly discriminated on the basis of this test.

A one-sample version of this index could be created by fixing the
normal density as the null hypothesis, and investigating the power
of the one-sample Kolmogorov--Smirnov using logistic samples. As seen
in Table~\ref{tb:ks 1}, this test is considerably more powerful than
the two-sample one.
\begin{table}[b]
\tablewidth=6.5cm
\caption{Power of the Kolmogorov--Smirnov test (one-sample method) to
detect the difference between normal and logistic distributions at
selected sample sizes}\label{tb:ks 1}
\begin{tabular*}{\tablewidth}{@{\extracolsep{\fill}}rc@{}}
\hline
\multicolumn{1}{@{}l}{$\bolds{m}$} & \textbf{Rejection proportion}\\
\hline
100 & 0.126 \\
400 & 0.435 \\
450 & 0.479 \\
$\hat{N}^*=485$ & 0.500 \\
500 & 0.518 \\
1000 & 0.824 \\
2500 & 1.000 \\
\hline
\end{tabular*}
\end{table}

\begin{table*}
\caption{Simulated power of normality test for finite population from logistic} \label{log tb}
\begin{tabular*}{\textwidth}{@{\extracolsep{\fill}}r cc cc c d{3.2}@{}}
\hline
& \multicolumn{2}{c}{\textbf{Bootstrap}} & \multicolumn{2}{c}{\textbf{Subsampling}} & & \multicolumn{1}{c@{}}{\textbf{Subsampling}}
\\
\ccline{2-3,4-5,7-7}
 \\[-6pt]
\multicolumn{1}{@{}l}{$\bolds{n}$} & \textbf{Mean} & \textbf{Deviation} & \textbf{Mean} & \textbf{Deviation} & $\bolds{\phi^{-1}= \frac{n}{m}}$ & \multicolumn{1}{c@{}}{\textbf{EISS}} \\
\hline
1000 & 0.659& 0.1753 & 0.493 & 0.2472 & \hspace*{4pt}$\frac{1000}{485}=2.06$ & 4.14
\\[2pt]
10,000 & 0.517& 0.0710 & 0.499 & 0.0735 & $\frac{10{,}000}{485}=20.6$ & 44.3
\\[2pt]
100,000 & 0.503& 0.0278 & 0.501 & 0.0252 & \hspace*{1pt}$\frac{100{,}000}{485}=206.2$ & 393.7 \\
\hline
\end{tabular*}%
\end{table*}

Note that this analysis also shows that $N^*$, when the model is
normal and the true distribution is logistic, is about 485, and so
logistic samples are closer to normality than is the blood pressure
data set.

Finally, we use this example to compare the bias and deviation of
$\hat{N}^*$ when estimated by bootstrap simulation with $\hat{N}^*$
when estimated by subsampling simulation. Consider a large data set
of size $n$ from the logistic distribution. We let $m$ be fixed and
simulate the powers of the one-sample Kolmogorov test for normality by
bootstrapping and by subsampling. We take 500 data sets from
logistic distribution at each size $n$. The simulated average and
standard deviation of power are in Table~\ref{log tb} for $m=485$
and $n= 1000$, 10,000 and 100,000.

The true power for the infinite population is approximately 0.5. The
results show that as the empirical data size $n$ gets larger and
larger, the simulated power gets closer and closer to the true
value. Although the standard deviations are almost the same for
the bootstrap method and the subsampling method, the simulated power by
bootstrap is much more biased for small $n$. With the bootstrap
method, sample size $485$ is estimated to have 0.66 power when $n$
is 1000. That again indicates that estimation of $N^*$ by bootstrap
tends to have a downward bias.

The reader should note the large standard deviation when $n=1000$.
The last two columns will be discussed later in the context of
understanding how well one can estimate power nonparametrically.

\section{Asymptotic Issues in Power Estimation}\label{sec4}

In this section we examine the asymptotic properties, as
$n\rightarrow \infty$, when one estimates the power curve
$\beta_{\tau }(m)$ by subsampling or bootstrapping.

Suppose our test statistic is $T_{n}=T_{n}(x_{1},\ldots,x_{n})$,
symmetric in
its arguments. Suppose our test procedure is to reject $H_{0}$ when
$\{T_{n}(X_{1},\ldots,X_{n})>c_{\alpha }\}$, where $c_{\alpha }$ is an
asymptotic critical value for the test. The object of interest is
\[
\beta (m)=P_{\tau }\{T_{m}(X_{1},\ldots,X_{n})>c_{\alpha }\}.
\]
When the null hypothesis is true (i.e., includes $\tau $), we have
$P_{\tau }\{T_{n}(X_{1},\ldots,X_{n})>c_{\alpha }\}\rightarrow \alpha $
as $n\rightarrow \infty $.

We will derive asymptotic results for subsampling based estimation
of $\beta (m)$, with side notes on the effect of using bootstrap
sampling instead. Notice that
$I\{T_{m}(X_{s_{1}},\ldots,X_{s_{m}})>c_{\alpha }\}$, for any set of
distinct integers $a_{1},\ldots,a_{m}$, is an unbiased estimator of
$\beta (m)$. Let $S=$ $\{s_{1},\ldots,s_{m}\}$ be a subset of $m$
distinct integers sampled from $\{1,\ldots,n\}$, and let
$X_{S}=(X_{s_{1}},\ldots,X_{s_{m}})$. Finally, let
$K_{m}(X_{S})=I\{T_{m}(X_{a_{1}},\ldots,\break X_{a_{m}})>c_{\alpha }\}$. We
can construct a $U$-statistic estimator of $\beta (m)$ by
\[
U_{\mathrm{comp}}(X)=\frac{1}{{n\choose m}}\sum_{S\in \mathbb{S}}K_{m}(X_{S}),
\]%
where $\mathbb{S}$ is the set of all distinct subsets of
$\{1,\ldots,n\}$ of size $m$. We can also write this as an expectation:%
\begin{equation}\label{complete u statistic}
U_{\mathrm{comp}}(X)=E  [ K_m(X_{S})|X_1, \ldots, X_n  ].
\end{equation}%
Here the expectation is over samples of $m$ integers without
replacement from $\{1,\ldots,n\}$, with $X=(X_{1},\break\ldots,X_{n})$ fixed.

We will call this the complete $U$-statistic; in practice, we are
unlikely to use it because of the ${n\choose m}$ calculations
required. The approximation we consider will replace this exact
expectation with a subsampling estimator created by randomly
sampling $S$. Another possible computational shortcut would be to
use a statistical design for the selection of a subset of $S$ (Blom, \citeyear{blom1976}).
We will focus here on the properties of $U_{\mathrm{comp}}$ itself,
corresponding to an ideal infinite subsampling scheme. In this
setting, we can think of the estimator obtained by bootstrap
subsampling as being the corresponding $V$-statistic estimator of
$\beta$.

\subsection{Fixed $m$ Asymptotics}

We can now make some observations about the consistency of this form
of
estimation. The answer depends on the asymptotic setting. If we assume that $%
m$ is held fixed as $n\rightarrow \infty $, \textit{fixed m
asymptotics, }then we can apply the following standard $U$-statistic
theory, and obtain consistency and asymptotic normality for the
estimation of $\beta (m)\ $as follows.

The exact and asymptotic variance of $U_{\mathrm{comp}}$ is described in
Theorem~\ref{thm u} (Lehmann, \citeyear{lehm1999}).
\begin{theorem}\label{thm u}
If $\operatorname{Var} [ K_m(x_1, \ldots, x_i, X_{i+1}, \ldots,\break X_{m})  ]
= \sigma^2_i$, then:
\begin{longlist}[(1)]
\item[(1)] The variance of the $U$-statistic is equal to
\[
\operatorname{Var}(U_{\mathrm{comp}}) = \sum_{i=1}^{m} \pmatrix{m \cr i} \pmatrix{n-m \cr m-i}\sigma^2_i \Big/\pmatrix{n \cr m}.
\]
\item[(2)] If $\sigma^2_1 > 0$ and $\sigma^2_i < \infty$ for all
$i=1,\ldots,m$, then
\[
\operatorname{Var}\bigl(\sqrt{n} U_{\mathrm{comp}}\bigr) \rightarrow m^2 \sigma^2_1.
\]
\end{longlist}
\end{theorem}

Theorem~\ref{thm u2} gives the asymptotic normal property of
$U_{\mathrm{comp}}$.
\begin{theorem} \label{thm u2}
\textup{(1)} If $0<\sigma^2_1<\infty$, then as $n \rightarrow \infty$,
\[ \sqrt{n} (U_{\mathrm{comp}}-\beta) \stackrel{d}  {\to} N(0, m^2 \sigma^2_1); \]

\textup{(2)} If $\sigma^2_i < \infty$ for all
$i=1, \ldots,m$, then
\[ \frac{U_{\mathrm{comp}}-\beta}{\sqrt{\operatorname{Var}(U_{\mathrm{comp}})}} \stackrel{d}  {\to} N(0,1). \]
\end{theorem}

Because for us $K_m$ is an indicator function, the condition that
$\sigma^2_i < \infty$ for all $i$ is obviously satisfied. When $m$
is fixed, these limiting distribution results hold for the bootstrap
estimator of $\beta(m)$ because it is the corresponding
$V$-statistic.

\subsection{Fixed Sampling Ratio Asymptotics}

Unfortunately using fixed $m$  asymptotics is incredibly optimistic
in our setting, as we wish to be able to estimate $\beta (m)$ for
$m$ as close to $n$ as possible. The more realistic asymptotics we
will use to study this case will consider sequences in $n$ in which
$m=m_{n}$ is some fixed fraction $\phi $ of $n$, which we call
\textit{fixed ratio asymptotics. }In this setting the target value
$\beta _{\tau }(m_{n})$ will be changing in $n$, going to $1$, and
so we also need to consider local alternative sequences $\tau _{n}$.

To study this, we first derive some properties of
$\operatorname{Var}(U_{\mathrm{comp}}(m))$. For
any two independent samples $S_{1}$ and $S_{2}$ of size $m$ from $%
\{1,2,\ldots,n\}$, let 
$|S_{1}\cap S_{2}|=O(S_{1},S_{2})$ be
the number of common elements. We will call $O(S_{1},S_{2})$ the \textit{%
sample overlap}. It has a hypergeometric distribution, so it is an
elementary calculation to show that $E\{O(S_{1},S_{2})\}/m=m/n=\phi
$. That is, the sampling fraction $\phi $ is also the mean
fractional overlap
between subsamples. We can then write%
\begin{eqnarray}\label{variance_formula}
\hspace*{15pt}&&E(U_{\mathrm{comp}}^{2}(m))\nonumber
\\
&&\quad=\sum_{k=0}^{m}E [ K_m(S_{1})K_m(S_{2})|O(S_{1},S_{2})=k ]
\\
&&\hspace*{13pt}\qquad{}\times\operatorname{Pr}  [ O(S_{1},S_{2})=k ].\nonumber
\end{eqnarray}%
As we will show below, the $U$-statistic can suffer a severe
degradation in variance, relative to the fixed $m$ asympotics, if
the mean overlap $\phi $ in the indices is too large. (Note that
$\phi $ goes to zero in fixed $m$ asymptotics, so the overlap mean
goes to zero.) As a way to measure the overlap effect, we define an
\textit{equivalent independent sample size} (EISS)
measure using the formula%
\[
\operatorname{Var}(U_{\mathrm{comp}}(m))=\frac{\operatorname{Var}(K_m(X_{1},\ldots,X_{m}))}{\mathit{EISS}}.
\]%
For our indicator kernel $K_m$ this gives the formula
\[
\operatorname{Var}(U_{\mathrm{comp}}(m))=\frac{\beta _{\tau }(m)(1-\beta _{\tau }(m))}{\mathit{EISS}}
\]%
and so we can think of EISS as being the sample size we would need
to conduct an IID experiment with equivalent accuracy in estimating
$\beta (m)$.

From a standard $U$-statistic inequality (Blom, \citeyear{blom1976}, page~574), we have
\begin{eqnarray}
\operatorname{Var}(U_{\mathrm{comp}})
&\leq& \frac{ \operatorname{Var}(I\{T_{m}(X(S))>c_{\alpha }\})}{n/m}\nonumber
\\[-8pt]\\[-8pt]
&=&\frac{\beta _{\tau }(m)(1-\beta _{\tau }(m))}{n/m}.\nonumber
\end{eqnarray}%
As a consequence, we are guaranteed consistent estimation of
$\beta_{\tau _{n}}(m_{n})$, along any sequence of alternatives
$\tau_{n}$, when $\phi =m_{n}/n$ goes to zero. We note that bootstrap
resampling does not have this strong guarantee of consistency, as
general results require $m^{2}/n$ to go to zero (Politis, Romano and Wolf, \citeyear{poliromawolf1999}).

This inequality also implies that $\mathit{EISS} \geq \phi ^{-1}=n/m$. That
is, $\phi ^{-1}$ gives us a lower bound for EISS for $\beta (m)$
inference. For example, a sampling fraction of $\phi =1/25$ is
guaranteed to provide at least as accurate an estimation of $p=\beta
(m)$ as would 25 draws from a Bernoulli distribution with success
probability $p$. As we will later see, this inequality can also be
thought of as an approximation when $\phi ^{-1}$ is small, helping
to give one the proper degree of pessimism about $N^{\ast } $
inference in this case.

\subsection{Local Alternatives: A Closer Look}

To more closely examine this approximation, we consider certain
local alternatives $\tau _{n}$ to the null hypothesis. We will
assume now that the test statistic at hand admits a standard local
asymptotic analysis under alternatives of the form $\tau
_{n}=F_{0}+n^{-1/2}cg(x)$, for fixed $g(x)$, positive $c$ and null
element $F_{0}$. In this setting one can typically show that $\beta
_{\tau _{n}}(n)\rightarrow \beta _{\mathrm{loc}}(c)$ as $n\rightarrow \infty
$, where the local alternative power curve $\beta _{\mathrm{loc}}(c)$ is a
continuous increasing function of $c$. For example, for Pearson's chi-square
test, the local analysis leads to a noncentral chi-square
distribution. (See Ferguson, \citeyear{ferg1996}, page~63.) To find the local power
along the sequence $\tau_n$ when a different sample size is used, say,
$m_{n}=\phi n$, we can rewrite the alternative as
\[
\tau _{n}=F_{0}+m_{n}^{-1/2}\phi ^{1/2}cg(x).
\]%
The sample size changes the scaling factor from $c$ to
$\phi^{1/2} c$. Hence, the asymptotic power approximation for samples of size $m_{n}$ from $%
\tau _{n}$ is $\beta _{\mathrm{loc}}(\phi ^{1/2}c)$. Assuming that $c$ is
chosen so that $\beta _{\mathrm{loc}}(c)>1/2$, there will be a fraction $\phi
_{0.5}$ such that $\beta _{\mathrm{loc}}(\phi _{0.5}^{1/2}c)=1/2$. That is, if
we choose $\phi=\phi_{0.5}$, we have $\beta _{\tau
_{n}}(m_{n})\rightarrow 0.5$, for $m_n=\phi_{0.5} n$. As a
consequence, the true $N^*$ value for the $\tau_n$ sequence grows
proportionally to $n$, namely, $\phi _{0.5}\times n$.

Since $\phi =m/n$ is fixed, our proceeding result about the consistency of $%
U_{\mathrm{comp}}$ is not operative. In fact, in local alternative settings,
the estimator is generally not consistent. However, it is possible
to obtain useful understanding of how the variance changes as a
function of $\phi$, and so examine its role
in estimation. Returning to the formula%
\begin{eqnarray*}
E(U_{\mathrm{comp}}^{2}(m))
&=&\sum_{k=0}^{m}E [ K_m(S_{1})K_m(S_{2})|O(S_{1},S_{2})=k]
\\
&&\hspace*{13pt}{}\times \operatorname{Pr}  [ O(S_{1},S_{2})=k ],
\end{eqnarray*}
the second term on the right has the elementary calculation
\[
\operatorname{Pr}  [ O(S_{1},S_{2})=k ] =\frac{{m\choose k}{n-m\choose m-k}}{%
{n\choose m}}.
\]%
This hypergeometric distribution has mean $\phi_m =(m/n) \cdot m$
and variance bounded above by $m\phi (1-\phi )$, the corresponding binomial variance. Hence,\break $%
O(S_{1},S_{2})/m$, the fractional overlap, converges in probability
to $\phi $ in our asymptotic setting.

For this reason, it is reasonable to approximate the terms
\[
E [K_m(S_{1})K_m(S_{2})|O(S_{1},S_{2})=k_{n} ]
\]
along a sequence of $%
k$'s for which the samples have a fixed fractional overlap, say,
$k_{n}=am_{n}=a \phi n$, in order to approximate the important terms
in the variance.

Although such a task is dependent on the structure of the test
statistic, we think it is worthwhile to illustrate here how these
calculations could be carried out. We consider a test statistic
which is asymptotically chi-squared distributed, with degrees of
freedom $d$ under the null hypothesis, and is asymptotically
noncentral chi-squared, with noncentrality parameter $\delta $
under the local alternatives sequence.

If we let $G(t)=\operatorname{Pr} \{\chi _{d-1}^{2}>t\}$, then for fixed overlap fraction $%
a$, then under standard local asymptotic calculations,
\[
E [ K_m(S_{1})K_m(S_{2})|O(S_{1},S_{2})=k_{n} ] \rightarrow
A,
\]%
where $A$ can be calculated as the expectation of
\begin{eqnarray}\label{bigG}
&&G \Biggl( \frac{1}{1-a}c_{\alpha }\nonumber
\\
&&\hspace*{14pt}{}- \Biggl( X+\Biggl(\Biggl(Z\sqrt{\frac{a}{1-a}}
+\sqrt{\frac{1}{1-a}}\delta \Biggr)^{2}\nonumber
\\
&&\hspace*{127pt}{}+\frac{a}{1-a}W\Biggr)^{1/2} \Biggr) ^{2} \Biggr)
 \nonumber
\\[-8pt]\\[-8pt]
&&{}\times G \Biggl( \frac{1}{1-a}c_{\alpha }\nonumber
\\
&&\hspace*{28pt}{}- \Biggl( Y+\Biggl(\Biggl(Z\sqrt{\frac{a}{1-a}}
+\sqrt{\frac{1}{1-a}}\delta\Biggr)^{2}\nonumber
\\
&&\hspace*{140pt}{}+\frac{a}{1-a}W\Biggr)^{1/2} \Biggr) ^{2} \Biggr)\nonumber
\end{eqnarray}%
where $X,Y,Z$ are independent normal variables and $W$ is independently $%
\chi _{d-1}^{2}$.

In Table~\ref{local alt} we show some calculations from this formula for $d=25$, where $%
\delta $ is chosen as $3.67$ so as to obtain asymptotic power 0.5.
The critical value is $c_{0.05}=37.66$.


We note several features here. First, $\phi^{-1}$ is relatively
conservative, but for small values does provide the right caution.
Here $\phi^{-1}=10$ gives an EISS of 32.6, something like a bare
minimum needed for $N^*$ inference. If we compare this table with
the values from the simulation in Table~\ref{log tb}, we see that in the
latter, EISS was about $2 \times \phi^{-1}$ across a larger range of
sampling fractions, and so did not show the steady improvement found
in Table~\ref{local alt}.

\section{Credibility in Categorical Data Models}\label{sec5}

Our setting for analyzing the mathematical features of credibility
indices more carefully will be likelihood ratio tests in categorical
models.

\subsection{Asymptotic Approximations} \label{thetatau}

We derive two approximations to $N^*$ here, focusing on the
likelihood ratio test in multinomial models. Here the data will be
an IID sample from a multinomial distribution, as summarized by the
counts $n(t)$ in the cells $t=1,\ldots,T$. The cell proportions will be
denoted $d(t)=n(t)/n$, which represent the empirical distribution
$\mathbf{d} $ of the data. The model $\mathcal{M}$ will have elements
$F_{\theta}(t)$ representing a parametric model for the multinomial
cells---for example, a log-linear model. The testing statistic will
be the likelihood ratio, and we will assume that the test statistics
have the standard asymptotic chi-squared distributions under the
null models.

\begin{table}
\caption{Simulated EISS for various sampling fraction $\phi$}\label{local alt}
\begin{tabular*}{\columnwidth}{@{\extracolsep{\fill}}ld{2.1}ld{3.1}ld{3.1}@{}}
\hline
\multicolumn{1}{@{}l}{$\bolds{\phi^{-1}}$}
& \multicolumn{1}{c}{\textbf{EISS}}
&  \multicolumn{1}{c}{$\bolds{\phi^{-1}}$}
& \multicolumn{1}{c}{\textbf{EISS}}
& \multicolumn{1}{c}{$\bolds{\phi^{-1}}$}
& \multicolumn{1}{c@{}}{\textbf{EISS}} \\
\hline
\phantom{0}2   &   4.2     &   15  &   52.9    &   \phantom{0}50  &   231.6  \\
\phantom{0}3   &   7.4     &   20  &   74.6    &   \phantom{0}60  &   294.1  \\
\phantom{0}4   &   10.7    &   25  &   97.7    &   \phantom{0}75  &   398.0  \\
\phantom{0}5   &   14.1    &   30  &   122.0   &   \phantom{0}80  &   435.6  \\
10  &   32.6    &   40  &   174.3   &   100 &   601.7  \\
\hline
\end{tabular*}%
\end{table}

In this context we can derive a simple asymptotic version of the
testing index and show that it is proportional to a reciprocal
squared distance. This in turn leads to an elementary consistent
estimator of the asymptotic index. This estimator has two important
uses:  It can be used for a preliminary value of the index for
bootstrap or subsampling testing. It can also itself be bootstrapped
or subsampled, which then provides a simple way to assess the
variability of the estimated index.

The \textit{likelihood deviation} between a multinomial distribution $%
\mathbf{p}$ and a model element $\mathbf{F}_{\theta}$ is defined as
$L^{2}(p,F_{\theta})=\sum p(t)\log   (p(t)/F_{\theta}(t)
 )$. This is a version of the Kullback--Leibler distance; we
call it the likelihood deviation to clarify the asymmetric role of
$\mathbf{p}$ and $\mathbf{F}$. Technically it operates as a squared
distance, which is why we use the superscript 2. We also define the
likelihood deviation from a multinomial distribution $\mathbf{p}$ to the
model $\mathcal{M}$ to be
\begin{equation}\label{eq:l2 def}
L^{2}(\mathbf{p},\mathcal{M})=\inf_{\theta } L^{2}(\mathbf{p},\mathbf{F}%
_{\theta }).
\end{equation}%
For the true sample distribution $\bolds{\tau}$, if the infinum is
attained at a particular $\theta$, it will be denoted $%
\theta _{\tau }$, and the model element that approximates
$\bolds{\tau}$ is therefore denoted $\mathbf{F}_{\theta_{\tau}}$.

In the likelihood ratio test, one rejects the null hypothesis $H_{0}$: $%
\bolds{\tau}\in \mathcal{M}$ at asymptotic size $\alpha $, if the likelihood ratio test
statistic is large enough, that is,
\[
2nL^{2}(\mathbf{d},\mathcal{M})\geq \chi _{\mathit{df}}^{2}(\alpha ),
\]%
where $\chi _{\mathit{df}}^{2}(\alpha )$ is the upper $1-\alpha$ quantile of
chi-squared distribution with $\mathit{df}=$ the degrees of freedom. The
power of the test at sample size $n$ when $\mathbf{d}_{n}\sim
\bolds{\tau}\notin \mathcal{M}$ is
\[
P_{\tau } \{ 2nL^{2}(\mathbf{d}_{n},\mathcal{M})\geq \chi
_{\mathit{df}}^{2}(\alpha ) \}.
\]
Our goal is to determine the sample size $N^{\ast }$ at which the
testing power for the alternative $\bolds{\tau }\notin
\mathcal{M}$ is 0.5. That is,
\[
P_{\tau } \{ 2N^{\ast }L^{2}(\mathbf{d}_{N^{\ast
}},\mathcal{M})\geq \chi _{\mathit{df}}^{2}(\alpha ) \} =0.5.
\]%

Our first approximation to $N^*$ uses the fact that when the model
is false, the centered likelihood ratio statistic has,
asymptotically, a centered normal distribution. The approximation,
as derived in the \hyperref[append]{Appendix}, is
\begin{equation}\label{eq:nstar by L2}
N_{\mathrm{asy}}^{\ast }(\tau )=\frac{\chi _{\mathit{df}}^{2}(\alpha )}{2L^{2}(\tau,\mathcal{%
M})}.
\end{equation}%
Here our choice of the power $0.5$ greatly simplifies the expression.
Other choices for $N^*_{\beta}$ would depend on the limiting
variance for the normal distribution.

Our second approximation is a bit more sophisticated. We consider
local alternatives that approach the null as the sample size goes to
infinity. This gives a noncentral chi-square approximation:
\begin{equation}\label{eq:nstar chi}
N_{\mathrm{asy} 2}^{\ast }(\tau )=\frac{(\delta^*)^2}
{X^2(\tau,\mathcal{M})}.
\end{equation}%


In equation (\ref{eq:nstar chi}), $X^2(\tau, \mathcal{M})$ is the
Pearson chi-square distance, \[ X^2(\tau,F)= \sum
\frac{(\tau-F)^2}{F},\] and $(\delta^*)^2$ is the noncentrality
parameter that satisfies
\begin{equation}\label{delta}
P \{ \chi'^{ 2}_{\mathit{df}}  ( (\delta^*)^2  ) > \chi^2_{\mathit{df}} (\alpha) \}=0.5,
\end{equation}
where $\chi'^{ 2}_{\mathit{df}}(\delta^2)$ is a
noncentral $\chi^2$ distribution with degrees of freedom $\mathit{df}$ and
noncentrality parameter $(\delta^*)^2$. One can generalize this
approximation by changing the right-hand side of (\ref{delta}) to a
chosen  power level. See the \hyperref[append]{Appendix} for more details.

The second approximation should be more accurate than the first for
situations when $\tau $ is close to the model. Notice that both
approximations (\ref{eq:nstar by L2}) and (\ref{eq:nstar chi}) show
an inverse relationship to squared distance. Moreover, we can see
that $\alpha$ plays a role only in the numerator of the
approximation. Given two models with the same testing degrees of
freedom, the ratio of approximate $N^*$-values does not depend on
$\alpha$.

Another useful feature of $N_{\mathrm{asy}}^{\ast }$ arises in confidence
assessment. One could form asymptotic confidence intervals for
$N^{\ast }(\tau )$ by bootstrapping $\hat{N}^{\ast }$, but this
requires double bootstrapping, an expensive possibility. But
bootstrapping $N_{\mathrm{asy}}^{\ast }(\mathbf{d})$ is relatively inexpensive
and it can give a useful picture of the uncertainty involved. More
rigorous methods of using subsampling to estimate standard errors
are under investigation by the authors.


\subsection{Numerical Examples}

We next assess model credibility for the data in Tables~\ref{tb:3.1} and~\ref{tb:3.2}. Table~\ref{tb:3.1}, considered
earlier by Snee (\citeyear{snee1974}), is a $4\times 4$ table cross-classifying eye
color and hair color. The sample size $n=592$ is somewhat large, but
the table does have some small entries. The Pearson statistic for the independence model is $%
X^{2}=138.290$ on $9$ degrees of freedom, and the likelihood ratio statistic
is $L^{2}=146.444$. The model would be rejected on the basis of these
quantities.

\begin{table}[b]
\caption{Cross-classification of eye color and hair color (size $n=592$)}\label{tb:3.1}
\begin{tabular*}{\columnwidth}{@{\extracolsep{\fill}}ld{2.0}d{3.0}d{2.0}d{2.0}@{}}
\hline
 & \multicolumn{4}{c@{}}{\textbf{Hair color}} \\
  \ccline{2-5} \\[-6pt]
\textbf{Eye color}
& \multicolumn{1}{c}{\textbf{Black}}
& \multicolumn{1}{c}{\textbf{Brunette}}
& \multicolumn{1}{c}{\textbf{Red}}
& \multicolumn{1}{c@{}}{\textbf{Blonde}} \\ \hline
Brown & 68 & 119 & 26 & 7 \\
Blue & 20 & 84 & 17 & 94 \\
Hazel & 15 & 54 & 14 & 10 \\
Green & 5 & 29 & 14 & 16 \\ \hline
\end{tabular*}
\end{table}

We tested the independence model for the data in Table~\ref{tb:3.1}, where the
degrees of freedom are 9. We then apply the two approximations, (\ref%
{eq:nstar by L2}) and (\ref{eq:nstar chi}), to obtain the starting
value for $N^{\ast }(\mathbf{d})$, which are $N_{\mathrm{asy}}^{\ast }(\mathbf{d})$
$=34$ and $N_{\mathrm{asy} 2}^{\ast }(\mathbf{d})$ $=37$.



We further refine the preliminary value by bootstrap. Given the target size $%
\alpha =0.05$, we took various sample sizes $m$, then generated $%
B=1000$ bootstrap samples $\mathbf{d}_{b}^{\ast }$ from $\operatorname{Multinomial}(m,\mathbf{d})$, with margins not fixed. We then conducted the
size $\alpha $ likelihood
ratio test, and recorded the fraction of rejections,
$\# \{ 2nL^{2}(\mathbf{d}_{b}^{\ast },\mathcal{M})\geq \chi _{\mathit{df}}^{2}(\alpha) \} /B$.
The estimate of $N^{\ast }(\tau )$, $N^{\ast}(\mathbf{d)}$,
would be that sample size that gives rejection fraction
$50\%$. See Table~\ref{tb:9.1} for the numbers, as well as a comparison of
bootstrap and subsampling in this example.

In this case $N^{\ast }(\mathbf{d})=32$, which is very close to the
first asymptotic value of $34$. A 95\% bootstrap interval for
$N_{\mathrm{asy}}^{\ast }(\tau )$ was found to be $(25,43)$. Note that
$\phi^{-1}=592/32=18.5$, suggesting that inference about $N^*$ is
reasonable.

Diaconis and Efron (\citeyear{diacefro1985}), in addressing the same problem posed by this
paper, suggested a different way of generating an assessment of this
particular data set. They compared the observed $X^{2}$-value with those of
all possible $4\times 4$ tables with $n=592$. They found that, among all $%
4\times 4$ tables with $n=592$ (margins not fixed), approximately $10\%$
have $X^{2}$-values less than $138.29$. They concluded that the given $%
4\times 4$ table does not lie particularly close to independence.

Our second example, Table~\ref{tb:3.2}, originally published in Cram\'{e}r
(\citeyear{Cramer1946}), is a $5\times 4$ table cross-classifying number of children by
annual income levels. The sample size is $n=25{,}263$, which is very large.
The goodness-of-fit statistics are $X^{2}=568.566$ and $L^{2}=569.420$ on 12
degrees of freedom. The $\chi ^{2}$-statistics have extremely small $p$%
-values, leading to rejection using the conventional criteria.

Diaconis and Efron (\citeyear{diacefro1985}) used this example as well. They found
that, among all $5\times 4$ tables with $n=25{,}263$ (margins not
fixed), the proportion of those having $X^{2}$ less than $568.576$
is $2.1\times 10^{-7}$. They concluded that the observed table is
extremely close to independence, which is dramatically opposite from
the conclusion drawn from the $\chi ^{2}$-values.
\begin{table}
\caption{Cross-classification of number of children by annual income\break(size $n=25{,}263$)}\label{tb:3.2}
\begin{tabular*}{\columnwidth}{@{\extracolsep{\fill}}ld{4.0}d{4.0}d{4.0}d{4.0}@{}}
\hline
 & \multicolumn{4}{c@{}}{\textbf{Annual income}}
\\[-6pt]
& \multicolumn{4}{c@{}}{\textbf{\hrulefill}} \\
\textbf{No. of children}
& \multicolumn{1}{c}{\textbf{0--1}}
& \multicolumn{1}{c}{\textbf{1--2}}
& \multicolumn{1}{c}{\textbf{2--3}}
& \multicolumn{1}{c@{}}{$\bolds{3+}$} \\
\hline
\phantom{$+$}0 & 2161 & 3577 & 2184 & 1636 \\
\phantom{$+$}1 & 2755 & 5081 & 2222 & 1052 \\
\phantom{$+$}2 & 936 & 1753 & 640 & 306 \\
\phantom{$+$}3 & 225 & 419 & 96 & 38 \\
4$+$ & 39 & 98 & 31 & 14 \\ \hline
\end{tabular*}
\end{table}

\begin{table*}
\tablewidth=12cm
\caption{Summary of sample sizes and the corresponding power for data in
Tables~\protect\ref{tb:3.1} and \protect\ref{tb:3.2}}\label{tb:9.1}
\begin{tabular*}{\tablewidth}{@{\extracolsep{4in minus 4in}}rccc@{\hspace*{-10pt}} lcc@{}}
\hline
\multicolumn{3}{@{}c}{\textbf{Power for Table~\ref{tb:3.1}}}
&& \multicolumn{3}{c@{}}{\textbf{Power for Table \ref{tb:3.2}}} \\
\ccline{1-3,5-7} \\[-6pt]
\multicolumn{1}{@{}l}{$\bolds{m}$} & \textbf{Bootstrap} &
\textbf{Subsampling}&
& $\bolds{m}$ & \textbf{Bootstrap} & \textbf{Subsampling} \\
\hline
34 & 0.676 & 0.568                 &    & \phantom{$\hat{N}^*=\ $}470 &0.578 & 0.548 \\
$\hat{N}^*= 32$ &0.505 &0.497      &    & \phantom{$\hat{N}^*=\ $}450 &0.544 & 0.529 \\
31 & 0.512 & 0.484                 &    & \phantom{$\hat{N}^*=\ $}430 &0.505 & 0.507 \\
30 &0.481 & 0.474                  &    & $\hat{N}^*= 425$ &0.495 & 0.500 \\
29 &0.480 & 0.467                  &    & \phantom{$\hat{N}^*=\ $}400 &0.482 & 0.479 \\
\hline
\end{tabular*}%
\end{table*}

The credibility index for Table~\ref{tb:3.2} was calculated as
follows. The starting estimate value of $N_{\mathrm{asy}}^{\ast }(\mathbf{d})$
for the data in Table~\ref{tb:3.2} was $470$ and its bootstrap range
was $(386,548)$, while $N_{\mathrm{asy} 2}^{\ast }(\mathbf{d})=439$. We refined
the estimate to $N^{\ast }(\mathbf{d})=425$ using the bootstrap
procedure (margins not fixed). Here the closeness of the model and
sample explains why $N_{\mathrm{asy} 2}^*$ worked better as a bootstrap
starting value. Note that $\phi^{-1}=25{,}263/425=59.4$, suggesting
that inference on $N^*$ is reasonable. See Table~\ref{tb:9.1} for more details.

It is clear that Table~\ref{tb:3.2} lies much closer to the
independence model than Table~\ref{tb:3.1}. Using the credibility
index as a guide, we would say that the row-column independence
model is credible only for
samples of size $N=32$ or smaller for the population represented by
Table~\ref{tb:3.1}. Table~\ref{tb:3.2} is credible for samples that are
more than ten times as large.

The magnitude of the ratio for the Efron--Diaconis statistics is on a
completely different scale, being $4.8 \times 10^{5}$. Of course,
the statistics involved are quite different in interpretation. The
Efron--Diaconis statistic and our index are not asking the usual
questions for contingency tables. The Efron--Diaconis statistic seems
to ask ``is this table surprisingly close to independence?'' It is
calculated by assuming that prior to data collection, every possible
table of that sample size was equally likely.
We ask instead, ``does this table come from a population that
generates samples that look independent, even for large $n?$''

\section{Discussion}\label{sec6}

The statistical community is currently facing an enormous challenge
(and opportunity) that arises from the new data generating capacity
of science and engineering. 
This paper has been concerned with the question: ``How should we reconcile our
parametric modeling tools with the fact that in a truly large data
set, parametric models are either clearly false or are too complex
to be concise descriptors of the key data features?'' We have tackled
one small part of this problem, assessing the quality of a model's
fit while assuming it is false. We have done so by modifying
hypothesis testing methods so that they can be used from a model
false perspective.

If model credibility indices are a good idea, then many questions
remain. For example, can we design the test procedures, and the corresponding $%
N^{\ast }$ values, that would reassure us about the robustness of
using a standard model-based statistical procedure? Is there a good
way to use $N^*$ quantifying, in an absolute sense, what it means for a
model to be a surprisingly good fit to a set of data, as in saying
that a data set is ``highly normal''? The theoretical development of
this idea might involve comparison of the credibility of the chosen
model with a randomly selected model with the same number of
parameters.

Another issue regards the comparison of $N^*$-values in models
across differing numbers of parameters. One possibility is to create
an index that adjusts for the number of parameters, such as $N^{\ast
}/(\#$ parameters). The form of such an index then could depend on
how we might ``expect'' $N^{\ast }$ to grow when the number of
parameters grows, given a sequence of arbitrary models.

Although we recognize that the ideas presented here are only a
beginning, we hope the reader has found them to be stimulating.

\begin{appendix}
\section*{Appendix: Two Approximations to $N^*$}\label{append}

\subsection{Approximation Through Normal Distribution}

We can obtain a quick-and-dirty approximation using the fact
that---when the model is false---the centered likelihood ratio
statistic has, asymptotically, a centered normal distribution.
%
%
\begin{lemma}
If $\{n(t)\}$ are a multinomial sample of size $n$ from a fixed
distribution $\tau $ not in $\mathcal{M}$, then as $n\rightarrow
\infty$,
\[
\sqrt{n} \bigl( L^{2}(\mathbf{d}_{n},\mathcal{M})-L^{2}(\bolds{\tau }%
,\mathcal{M}) \bigr) \longrightarrow N(0,\sigma ^{2}),
\]%
provided that the asymptotic variance $\sigma ^{2}$
is not zero or infinity.
\end{lemma}

The lemma is just the maximum likelihood within von Mises' framework
(Serfling, \citeyear{Ser80}, page~211). Freitag and Munk (\citeyear{FreitagMunk2005}) have a
bootstrap variant, which is an interesting extension of the lemma.

Note that this lemma applies to bootstrap sampling from the
empirical distribution $d(t)$ (treating it as $\tau )$ whenever the
data $d(t)$ is not perfectly fit by the model. Now the value of $N$
that we seek satisfies
\begin{eqnarray*}
&&P \biggl\{ \sqrt{N}L^{2}(\mathbf{d}_{N},\mathcal{M})-\sqrt{N}L^{2}(\tau,\mathcal{M})
\\
&&\hspace*{4pt}\quad\geq \frac{1}{2\sqrt{N}}\chi _{\mathit{df}}^{2}(\alpha )-\sqrt{N}%
L^{2}(\tau,\mathcal{M}) \biggr\} =0.5.
\end{eqnarray*}
Since the left-hand term is asymptotically normal with mean zero,
this suggests that we need $N$ to solve
\[
\frac{1}{2\sqrt{N}}\chi _{\mathit{df}}^{2}(\alpha )-\sqrt{N}L^{2}(\tau
,\mathcal{M})=0.
\]%
Note that this calculation is independent of the unknown $\sigma
^{2}$ due to the choice of power $0.50$. It gives us the
approximation
\begin{equation}\label{eq:nstar1}
N_{\mathrm{asy}}^{\ast }(\tau )=\frac{\chi _{\mathit{df}}^{2}(\alpha )}{2L^{2}(\tau,\mathcal{%
M})}.
\end{equation}%
Thus, the asymptotic version of $N^{\ast }$ is inversely proportional
to the squared likelihood deviation.

Of course, our argument was somewhat specious: one cannot
simultaneously let $N$ go to infinity and solve for finite $N$.
Regardless, $N_{\mathrm{asy}}^{\ast}$ provides an elementary and useful
approximation to the index $N^{\ast }$, both its theoretical value
(sampling under $\tau $) and the estimator (sampling under
$\mathbf{d}$).

\subsection{Second Approximation to $N^*$ Using Noncentral
Chi-Square Distribution}

One could construct more sophisticated asymptotic approximations of
$N^*$. One method would be based on using ``local
alternatives''; that is, based on letting the alternatives approach
the null, as $n\rightarrow \infty $, obtaining noncentral
chi-square approximations.

We imagine a sequence of true alternatives with
$\tau_{m}=(1-m^{-1/2} ) F + m^{-1/2} g$, where $F$ is a model
element and $g$ is some fixed alternative not depending on $m$.
Therefore, the likelihood ratio test statistics $2 m L^2(d_{m},F)
\longrightarrow \chi'^{ 2}_{\mathit{df}}(\delta^2)$ as $ m \rightarrow
\infty$ under $\tau_{m}$, where $\delta^2 = X^2(g,F)$, the Pearson
chi-squared distance, $\sum (g-F)^2/F$, and $\chi'^{ 2}_{\mathit{df}}(\delta^2)$ is a noncentral chi-square distribution with
degrees of freedom $\mathit{df}$ and noncentrality parameter $\delta^2$
(Agresti, \citeyear{Agresti2002}).

Therefore, one can obtain the power as a function of $m$ at a fixed
$g$, based on the sequence of $\tau_{m}$. However, what we want is
the power at a particular $\tau$, which we can approximate by
inventing a different $g$ for each $m$. At the targeted $ m$,
\[
\tau = \tau_{ m} = (1-m^{-1/2}) F + m^{-1/2} g_{m}
\]
implies
\[
g_{m} = F + m^{1/2} (\tau - F).
\]

This gives the corresponding noncentrality parameter
\[
\delta^2 = \sum \frac{(g_{m}-F)^2}{F} =m X^2(\tau,F).
\]

We then get the power at $\tau$ for large $n$ being approximately
\[
P \{ \chi'^{ 2}_{\mathit{df}}(\delta^2) > \chi^2_{\mathit{df}}(\alpha)   \}.
\]
One can find the noncentrality parameter $(\delta^*)^2(\mathit{df})$ such
that
\[
P \{ \chi'^{ 2}_{\mathit{df}}((\delta^*)^2) > \chi^2_{\mathit{df}}(\alpha)   \}
=0.5,
\]
then $N^*$ can be approximated by
\begin{equation}  \label{eq:nstar2}
N^*_{\mathrm{asy} 2} = \frac{(\delta^*)^2(\mathit{df})}{X^2(\tau,F)}.
\end{equation}
\end{appendix}

\section*{Acknowledgment}
Supported in part by NSF Award DMS-04-05637.

\end{document}